\documentclass{jfm}
\usepackage{graphicx}
\usepackage{epstopdf, epsfig}

\shorttitle{Draft version \today}

\title{Relaminarising pipe flow by wall movement}
\author{D. Scarselli\aff{1}\corresp{\email{dscarsel@ist.ac.at}}, J.\ K\"{u}hnen\aff{1}, \and B. Hof\aff{1}}
\affiliation{\aff{1} Institute of Science and Technology Austria, Am Campus 1, A-3400 Klosterneuburg}

\graphicspath{{./figures/}}

\begin{document}

\maketitle

\begin{abstract}
Following the recent observation that turbulent pipe flow can be relaminarised by a relatively simple modification of the mean velocity profile, we here carry out a quantitative experimental investigation of this phenomenon. Our study confirms that a flat velocity profile leads to a collapse of turbulence and in order to achieve the blunted profile shape, we employ a moving pipe segment that is briefly and rapidly shifted in the streamwise direction. The relaminarisation threshold and the minimum shift length and speeds are determined as a function of Reynolds number. Although turbulence is still active after the acceleration phase, the modulated profile possesses a severely decreased lift-up potential as measured by transient growth. As shown, this results in an exponential decay of fluctuations and the flow relaminarises. While this method can be easily applied at low to moderate flow speeds, the minimum streamwise length over which the acceleration needs to act increases linearly with the Reynolds number.  
\end{abstract}

\begin{keywords}
Authors should not enter keywords on the manuscript, as these must be chosen by the author during the online submission process and will then be added during the typesetting process (see http://journals.cambridge.org/data/\linebreak[3]relatedlink/jfm-\linebreak[3]keywords.pdf for the full list)
\end{keywords}

\section{Introduction}

Techniques for relaminarisation of turbulent pipe flow are alluring mainly for two reasons. Firstly, from a technological point of view, laminar pipe flow is optimal in terms of net driving power in a controlled scenario \citep{Fukagata2009}, thus allowing in theory huge energy savings in pipeline systems. Secondly, a successful control of turbulence may provide better understanding and shed light on the dynamics on the phenomena involving production and dissipation of turbulence.

Several experimental investigations of relaminarizing pipe and channel flows have been reviewed by \citet{Sreenivasan1982}. However, the general experimental arrangement in the examples given involves a decrease in Reynolds number \citep[see e.g.][]{Sibulkin1962,Narayanan1968,Selvam2015}. Occasional evidence of relaminarisation not determined by dissipation and the Reynolds number has been found when a turbulent flow is subject to effects of acceleration, suction, blowing, magnetic fields, stratification, rotation, curvature or heating \citep{Sreenivasan1982}. In accelerated pipe flow, i.e. during and subsequent to a rapid increase of the flow rate of an initially turbulent flow, the flow has been observed to transiently visit a quasi-laminar state and undergo a process of transition that resembles the laminar-turbulent transition \citep[see e.g.][]{Lefebvre1989,Greenblatt1999,Greenblatt2004,He2013,He2015}. Temporary relaminarisation has also been reported for fluid injection through a porous wall segment in a pipe \citep{Pennell1972} .

\citet{hof-2010} introduced an alternative approach to suppress localised turbulent spots by reducing the inflection points in the mean axial velocity and more recently \citet{kuhnen-2018a,kuhnen-2018} have shown that a suitable modification of the mean velocity profile can lead to a complete collapse of turbulence, causing a turbulent flow to fully relaminarise. With the aid of numerical simulations and different experimental devices, the authors demonstrated that a plug-like, mean streamwise velocity profile has a reduced lift-up potential and leads to a complete collapse of turbulence. In particular, one technique was shown to laminarise the flow up to a Reynolds number of 40\,000 by strongly increasing the fluid velocity in the wall region. In these experiments, an initially turbulent pipe flow is perturbed by impulsively shifting a pipe segment that moves coaxially and relatively to the rest of the pipe. As a consequence, the fluid in contact with the moving segment is subject to a temporary modification of the boundary condition and experiences an injection of momentum into the near-wall region. After the wall stops, the perturbed flow undergoes a progressive laminarisation while being advected downstream.

In the present investigation we want to further explore the effect and possibilities of such a moving wall strategy in order to modify the streamwise velocity profile and control turbulent pipe flow. Different from \citet{kuhnen-2018}, we assess the circumstances under which turbulence fully decays by varying the wall velocity and shift length and we study the flow properties during and right after the wall movement up to a Reynolds number of 22\,000.

The idea of controlling the flow by a change of the boundary condition shares some aspects with drag reduction approaches in which a partial slip boundary condition is obtained by (super)hydrophobic walls and surfaces \citep[see e.g.][]{Watanabe1999,Joseph2005,Neto2005,Ou2005,Daniello2009,Rothstein2010,Yao2011,Lee2014,Saranadhi2016}. Slip on water repellent walls is usually realised in the range of nanometres. Only \citet{Saranadhi2016} report slip lengths of approx.\ 1\,mm by using active heating on a superhydrophobic surface to establish a stable vapour layer (Leidenfrost state), which is already two orders of magnitude larger than that achieved by the aforementioned authors. In the present study, however, during the perturbation phase we move the wall by amounts that range from centimetres to meters, in the order of tens of pipe diameters. The method used has also common features with the moving surface boundary-layer control used to delay flow separation through momentum injection \citep[see e.g.][]{Modi1997,Munshi1999}. In contrast to the more common ways of separation control in boundary layers (suction, blowing, vortex generators, turbulence promoters, etc.), these authors use moving surfaces such as a rotating cylinder at the leading edge of a flat plate or bluff bodies as momentum injecting elements.

The outline of the paper is as follows. In section \ref{sec.experiments} we describe the experimental facility and the measurement techniques employed. In section \ref{sec.results} we show the results of our investigation and in section \ref{sec.discussion} we put them in perspective by discussing the physics and the mechanisms at play during and after the wall shift. Finally, in \ref{sec.conclusions} we summarise our findings.

\section{Experiments}\label{sec.experiments}

\subsection{Wall movement apparatus}
\begin{figure}
  \centerline{\includegraphics[clip,width=0.92\textwidth,angle=0]{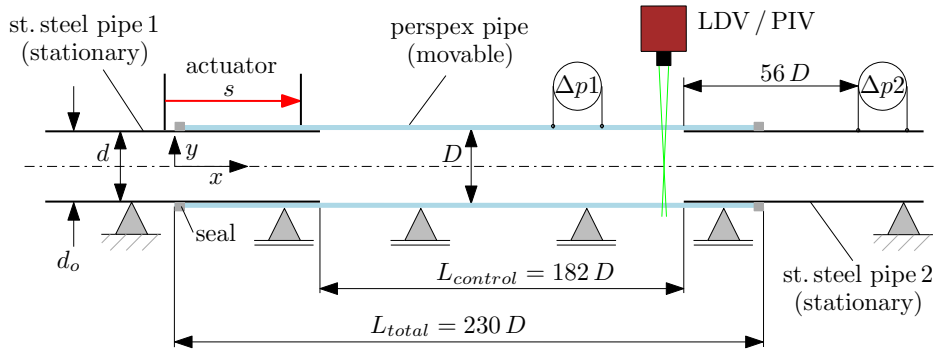}}
  \caption{\label{fig.sketch-setup_shift}Sketch of the experimental setup with a movable pipe segment. The movable pipe (perspex) is partly slipped over two stationary, very thin walled stainless steel pipes such that the perspex pipe overlaps the steel pipes at the upstream and downstream end. The perspex pipe is movable back and forth in the streamwise direction $z$ for an adjustable distance $s$ by means of a linear actuator. Drawing not to scale.}
\end{figure}

The setup consists of two consecutive straight stainless steel pipes (outer diameter $d_o = 25.4 \pm 0.13\,\textrm{mm}$, length 2$\,$m, wall thickness $0.4\pm0.04\,\textrm{mm}$) which are connected by a coaxial Perspex pipe with a slightly larger diameter ($D = 26\pm0.1\,\textrm{mm}$, length $L_{total}=230D\,$). This segment can be shifted back and forth along the axial direction at an adjustable speed for a prescribed distance. Figure \ref{fig.sketch-setup_shift} shows a sketch of the facility and indicates the arrangement of the measuring devices. The flow is driven by a constant pressure head.
The movable Perspex segment is partially slipped over the upstream and downstream steel pipes (respectively labelled 1 and 2 in figure \ref{fig.sketch-setup_shift}). The steel pipes are fixed to the base of the setup and support the Perspex pipe that is free to slide along the axial direction. Four polymer sleeve bearings (Igus) provide additional support to the moving section and help to prevent bending and vibrations during the movement. Two radial shaft seals are placed in the gap between the inner wall of the moving section and the outer one of the steel pipes to avoid leakage. The actual length of pipe that can be moved to modify the pipe wall velocity is $L_{control} = 182\,D$. Since the steel pipes have a smaller diameter than the control section, the flow experiences a small backward facing step at the end of the control section ($h=0.7\,\textrm{mm}, h/D=0.027$). We employ a linear actuator (toothed belt axis with roller guide driven by a servomotor, ELGA-TB-RF-70-1500-100H-P0, Festo; not shown in the figure) mounted beneath the pipe and clamped to the Perspex pipe to actuate the control. The actuator can move the perspex pipe for an adjustable distance (traverse path) $s\leq s_{max}=1.5 \,\textrm{m}$ at an adjustable velocity $U_w\leq U_{w, max}=5.5\,\textrm{m/s}$. The maximum possible acceleration is $a=50\,\textrm{m/}\textrm{s}^2$. Throughout the results presented in the present work the acceleration and deceleration ramps were kept constant to $|a|=10\,\textrm{m}/\textrm{s}^2$, unless otherwise specified.

In order to adjust the Reynolds number ($\Rey= U_b D/{\nu}$, where $\nu$ is the kinematic viscosity of the fluid and $U_b$ the bulk velocity) we regulate a valve located upstream of the test section in the supply pipe (not shown). The flow rate is monitored with an electromagnetic flow meter (ProcessMaster FXE4000, ABB) and the fluid temperature with a pt-100 resistance thermometer, both located in the feeding line. It is worth noticing that a change in the flow state (laminar or turbulent) in the test section does not appreciably affect $\Rey$, as the pressure drop difference along the main pipe (corresponding to $\approx 15$ mm of water at $\Rey=5000$) is negligible with respect to the total pressure head of $\approx 20\,\textrm{m}$. Hence, most of the pressure drop occurs across the regulation valve, which effectively keeps the mass flux (almost) constant throughout the measurements. The overall measurement accuracy is $\pm1\%$ for $\Rey$…

\subsection{Measurement techniques}

In order to investigate the flow behaviour during and after the wall shift we employ pressure drop measurements, particle image velocimetry (PIV) and laser Doppler velocimetry (LDV). Pressure drop measurements can easily detect the flow status (turbulent or laminar) after the wall stops and allow for a precise assessment of the skin friction. However, they fail to accurately capture the fast dynamics during the wall movement and immediately afterwards because of setup vibrations and the sensors slow response. The LDV system instead allows a more accurate description of the flow development throughout the experiment, although it does not provide information about the wall friction. The 2D PIV system offers a greater deal of data, but it is less suitable for investigations of a wide parameter space and it is hence used to study selected cases.

A first differential pressure sensor (DP\,45, Validyne) is mounted onto the movable perspex pipe $128\,D$ downstream of the beginning of the control section (distance measured when the actuator is not extended, $s=0$). The transducer is connected to two pressure taps of diameter 0.5\,mm, axial spacing 260\,mm and measures the pressure drop $\Delta p_1$. A second sensor (DP\,45, Validyne) is mounted on the steel pipe 2, $56\,D$ downstream the end of the control section. The taps have a diameter of 0.5\,mm and an axial spacing of 197\,mm and are associated to the pressure drop $\Delta p_2$. A great deal of care has been taken to stabilise the sensor housings and related wiring and piping during the impulsive pipe movement, especially to ensure repeatability. Overall, the measurement accuracy is $\pm 1.2\,\textrm{Pa}$.

At the downstream end of the movable perspex section, $5\,D$ upstream of steel pipe 2, the centreline velocity $U_c$ is measured by means of a one-component LDV system (TSI). Water is seeded with neutrally buoyant, hollow glass spheres of diameter $13\,\mu\textrm{m}$ (Sphericel, Potter Industries). The resulting average measurement rate is 20 Hz.

A 2D PIV system is set to monitor the flow along a longitudinal section of the movable pipe segment. The window is $1.5\,D$ long, centred in the same location as the LDV and passes through the centreline of the pipe. In order to decrease the distortion caused by refraction we enclose the aforementioned pipe segment in a rectangular, water filled Perspex box. A continuous laser (Fingco 532H-2W) illuminates the measurement plane with a sheet of light of nominal thickness $\approx 1\,\textrm{mm}$. PIV images are recorded with a high-speed camera (PhantomV10) mounted vertically above the water filled box. The flow is seeded with neutrally buoyant, hollow glass spheres of diameter $13\,\mu\textrm{m}$ (Sphericel, Potter Industries). The system is used to produce a 2D velocity field over a domain $1.5D\times D$ of resolution $87\times56$ vectors at a frequency of 50 Hz. The image post processing is carried out with the software Davis 8 (LaVision).

We also employ neutrally buoyant anisotropic particles (Mearlmaid Pearlessence) for visualising the flow state during and after the wall movement and for coarsely exploring the experimental parameter space. The particles have the form of elongated platelets that align with the local shear and possess high reflectivity allowing to observe flow structures \citep[see e.g.][]{Matisse1984}. An LED string is placed along the whole length of the movable Perspex pipe and illuminates the flow, enabling an easy detection of laminar and turbulent states both by the naked eye and camera. 

In the following we refer to a coordinate-system as indicated in figure \ref{fig.sketch-setup_shift}, where $x$ measures the axial direction along the flow and $y$ the wall-normal direction starting at the centreline. The respective Cartesian velocity components are $U$ and $V$.

\section{Results}\label{sec.results}
\begin{figure}
	\centerline{\includegraphics[width=\textwidth,angle=0]{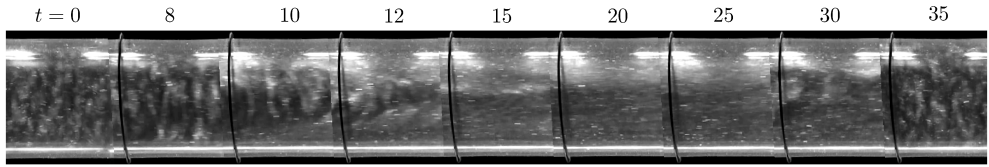}}
	\caption{Still pictures from the supplementary movie. Each frame shows a 1 $D$ long section of the movable perspex-pipe at $\Rey=5000$,  $s=9D$ and $U_w = U_b$. Time $t$ is measured in seconds. The wall is moved for $t=5-7$ s. The black ring around the pipe is used to highlight the movement of the wall.}
\label{fig.9stillframes}
\end{figure}

In an initial set of experiments  we set $\Rey = 5000$ and progressively increase the wall shift length $s$ while the wall velocity $U_w$ is kept equal to the bulk velocity $U_b$. The flow is monitored at the downstream end of the transparent pipe. As the shift length $s$ is increased, relaminarisation events begin to occur up to the point when at each actuation the flow consistently and repeatedly relaminarises for $s\gtrsim 8D$. It is important to note that the turbulence decay process takes place after the wall stops, so that the controlled patch of fluid is advected downstream while relaminarizing. Figure \ref{fig.9stillframes} shows still pictures from a typical run (supplementary movie available online). An initially fully turbulent flow at $\Rey=5000$ ($t<5\,$s) is subjected to an abrupt wall shift $(5<t<7\,$s) for a length $s=9D$ and wall speed $U_w = U_b$. After the wall stops, turbulent structures are still visible through the pipe (second panel in figure \ref{fig.9stillframes}). Nevertheless, they gradually decay and the flow reverts to the laminar state. Since laminarisation only occurs in the moved section, finally the laminarised flow patch is advected past and replaced by the upstream turbulent flow ($t>30\,$s).

\begin{figure}
\centering
\includegraphics[scale=.9]{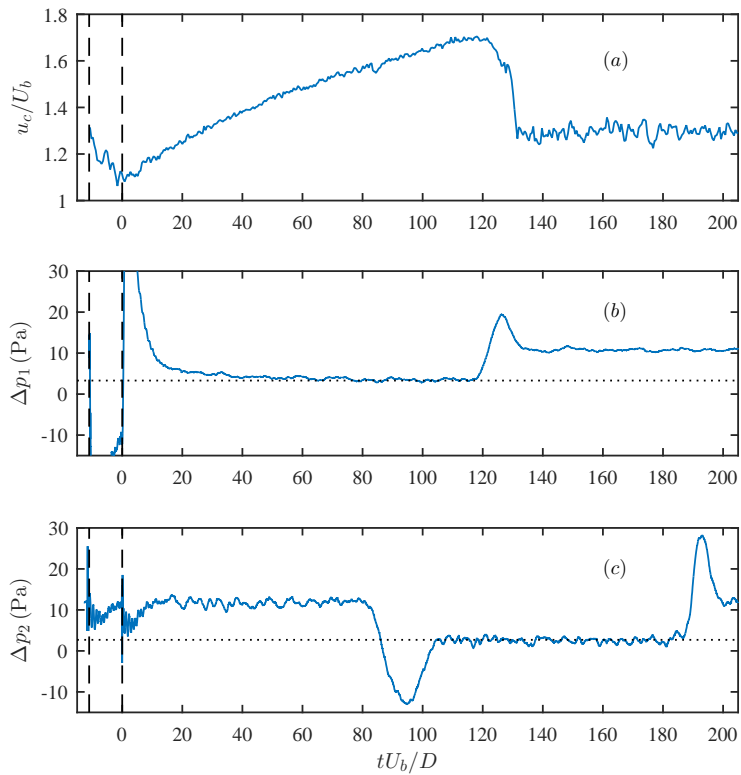}
\caption{Relaminarisation of a turbulent flow at $\Rey=8000$ with wall shift $s=12D$ and wall velocity $U_w = U_c$. (a) Centreline velocity over time measured by 2D PIV. Each point is obtained by averaging along the pipe axis for each frame. (b) Pressure drop $\Delta p_1$ over time read by the transducer mounted on the moving section. The off-scale portion of the signal reaches approximately -30 and 50 Pa. (c) Pressure drop $\Delta p_2$ over time read by the second transducer mounted downstream the moving section. The vertical dashed lines represent the first and last instant of wall motion. The dotted line marks the theoretical laminar pressure drop.}
\label{fig.u_dp1_dp2}
\end{figure}

\begin{figure}
\centering
\includegraphics[scale=.9]{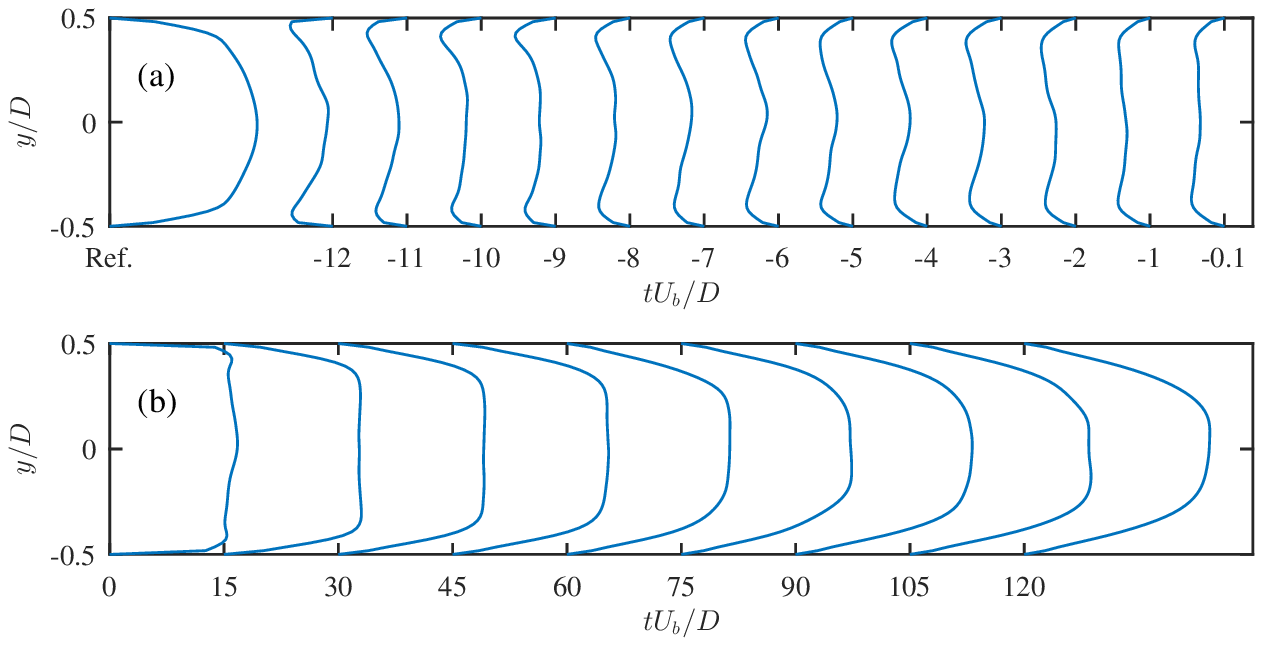}
\caption{Temporal evolution of the mean axial velocity during (a), the wall movement and (b), after it. The Reynolds number, wall shift length and velocity are $\Rey=8000$, $s=12D$ and $U_w = 1.3 U_b$, respectively. Time is measured in advective units starting from the end of the wall motion.}
\label{fig.re=8000_u_evolution_panel}
\end{figure}

Figure \ref{fig.u_dp1_dp2} shows the complete relaminarisation of a $\Rey=8000$ turbulent flow in terms of (a), average centreline velocity measured with PIV, (b) and (c), pressure drop $\Delta p_1$ and $\Delta p_2$, respectively.
The wall shift is $s=12D$ at $U_w = U_c \approx 1.3 U_b$, where $U_c$ is the average centreline velocity of the turbulent flow. The vertical dashed lines represent the first and last instant of wall motion. The dotted line marks the theoretical laminar pressure drop. During the shift the centreline velocity decreases steeply from $\approx 1.3U_b$ to $\approx 1.1 U_b$ (figure \ref{fig.re=8000_u_evolution_panel} (a), $-10<t<0$), whereas no reliable information is available from the pressure sensors (the signal goes off scale in figure \ref{fig.re=8000_u_evolution_panel} (b)), as a consequence of shaking and rapid wall shear change. Immediately after the wall stops the flow is steadily developing towards a parabolic profile, until it is advected past the measurement point. This is well captured in figure \ref{fig.u_dp1_dp2} (c), where at a slightly later time the downstream pressure taps record the passage of a short patch where the flow has fully relaminarised. The large overshoots visible in the pressure signals indicate the passage of the turbulent-laminar interface across the taps. It is worth noticing at this point that the laminarising patch stayed laminar even after passing over the step between the Perspex and steel pipes. 

A more precise insight into the dynamics during the wall motion is provided by figure \ref{fig.re=8000_u_evolution_panel} (a).
Here we show the temporal evolution of the axial velocity measured by 2D PIV. Each profile is averaged along the x axis of the PIV window and is labelled with the number of advective time units elapsed since the end of the wall movement. The profile marked as Ref. represents the undisturbed turbulent flow. At the beginning of the wall motion ($tU_b/D=-12$) the effect of the moving wall is confined to a region close to the wall. As time proceeds further, also the flow in the core region becomes progressively affected by the new boundary condition and the mean velocity assumes a flatter distribution. The flow development afterwards is shown in figure \ref{fig.re=8000_u_evolution_panel} (b). Immediately after the wall has stopped, the non-zero velocity boundary condition is  restored and the flows assumes a plug-like profile. From hereafter the flow develops towards parabolic, culminating with a centreline speed $U_c/U_b\approx 1.7$ before the laminar patch is advected downstream the observation window.

\begin{figure}
\centering
\includegraphics[scale=.9]{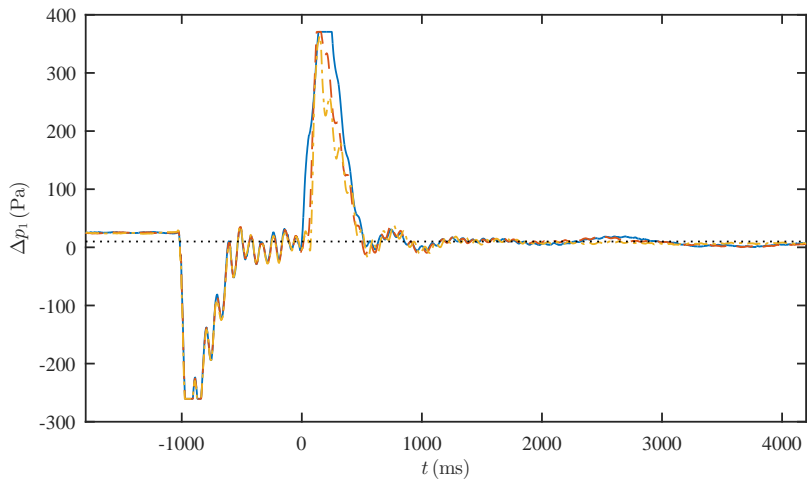}
\caption{Pressure signal $\Delta p_1$ of relaminarizing experiments at $\Rey=10\,000$, for $s=16D$ and $U_w = U_c \approx 1.3U_b$. Solid (blue online), dashed (red online) and dot-dashed (yellow online) lines correspond to accelerations $|a|=2,5,10\,\,\textrm{m}/\textrm{s}^2$. Each line is obtained by averaging three different runs. The dotted line represents the laminar pressure drop.}
\label{fig.10000_deceleration_comparison}
\end{figure}

The fate of the flow appears to depend only on the steady wall velocity and on the shift length during the wall motion phase. The acceleration ramp before and the deceleration ramp after do not appear to affect the results in the range of accelerations investigated, as shown in figure \ref{fig.10000_deceleration_comparison}. Here we compare the pressure signal $\Delta p_1$ for three relaminarising cases at $\Rey=10\,000$ with accelerations values of $|a|= 2,5,10\,\,\textrm{m}/\textrm{s}^2$ in solid (blue online), dashed (red online) and dot-dashed line (yellow online), respectively. Each signal is obtained by averaging three different runs to highlight coherent oscillations due to physical vibrations. Wall shift and wall velocity are $s=16D$ and $U_w = U_c \approx 1.3 U_b$, respectively. The dotted line represents the laminar pressure drop.

\begin{figure}
\centering
\includegraphics[scale=0.75]{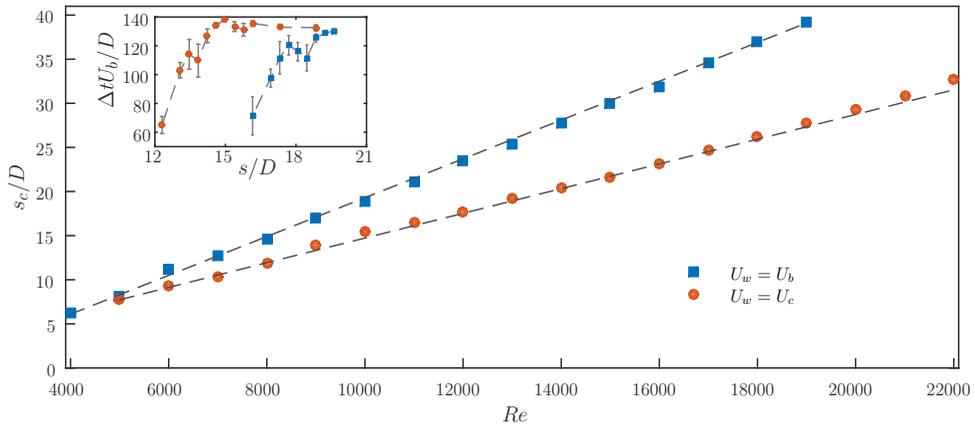}
\caption{Critical shift length $s_c$ for relaminarisation versus $\Rey$. Squares (blue online) and circles (red online) are datasets for $U_w = U_b$  and $U_w = U_c$, respectively. Dashed lines are a linear fit to the data. The inset shows the datasets used to assign the critical shift length for $\Rey=10\,000$. The average laminar patch duration $\Delta t$ is plotted against $s$. Error bars represent the standard deviation.}
\label{fig.criticalshiftlength_ubulk-vs-uopt_insert}
\end{figure}

We next explore how the shift length affects the laminarisation process. In these runs the flow status is monitored with LDV to detect laminar patches and measure their lengths. Figure \ref{fig.criticalshiftlength_ubulk-vs-uopt_insert} shows the minimum shift length required for laminarisation (hereafter referred to as the critical shift length $s_c$) versus Reynolds number for two wall velocities, $U_w = U_b$ (squares, blue online) and $U_w = U_c$ (circles, red online).
The dashed lines are a linear fit to the data. For $\Rey\lesssim 5000$ we do not observe any difference between the two velocities. Each point of the plot corresponds to a set of measurements where we increase $s$ and we keep the wall velocity constant. The inset of figure \ref{fig.criticalshiftlength_ubulk-vs-uopt_insert} shows one such dataset, where the mean length of the laminar patch $\Delta t$ is plotted versus $s$, for $U_w = U_b$ (squares, blue online) and $U_w = U_c$ (circles, red online). Error bars represent the standard deviation. The Reynolds number is $\Rey=10\,000$. As the shift $s$ is increased, the duration of the laminar flow increases and saturates. In order to determine the critical shift length we choose the first value of $s$ that lies in the saturated region and has a consistent repeatability (9 out of 10 runs relaminarising completely).

\begin{figure}
\centering
\includegraphics[scale=0.75]{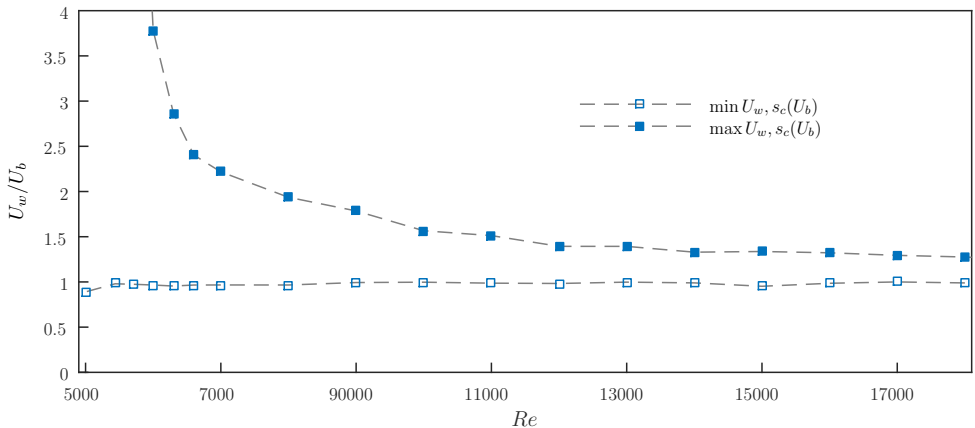}
\caption{Admissible values of wall velocity $U_w$ as a function of Reynolds number $\Rey$. The empty and full symbols represent respectively the minimum and maximum wall velocity necessary to relaminarise the flow with a critical shift length $s_c$ obtained from the data of figure \ref{fig.criticalshiftlength_ubulk-vs-uopt_insert} for $U_w=U_b$. At $\Rey=5000$ laminarisation was achieved up to $U_w = 40U_b$ (point not shown in figure).}
\label{fig.Shiftvel_min-max}
\end{figure}

Next, we explore the influence of the wall velocity as the shift length is held constant. For each $\Rey$ we pick the critical shift length $s_c$ for $U_w=U_b$ from figure \ref{fig.criticalshiftlength_ubulk-vs-uopt_insert}. For this shift length and $\Rey$ we then vary the wall speed and determine the speed range over which relaminarisation occurs. The minimum and maximum speed required is given respectively by the open and full symbols in figure \ref{fig.Shiftvel_min-max}. Each data point is found analogously to the search for the critical shift length.
As the Reynolds number increases, the allowable shift velocity range decreases rapidly while the minimum velocity seems to be independent of $\Rey$. Interestingly, at low Reynolds numbers ($\Rey<6000$) arbitrary large wall velocities lead to relaminarisation, for a shift equal to the critical value obtained with $U_w = U_b$. The maximum velocity tested was $U_w = 40U_b$ at $\Rey=5000$ (point not shown in the figure) and the flow fully relaminarised. Here we had to increase the acceleration to the maximum allowable value to reach the prescribed velocity.

\section{Discussion}\label{sec.discussion}

\begin{figure}
\centering
\includegraphics[scale=.9]{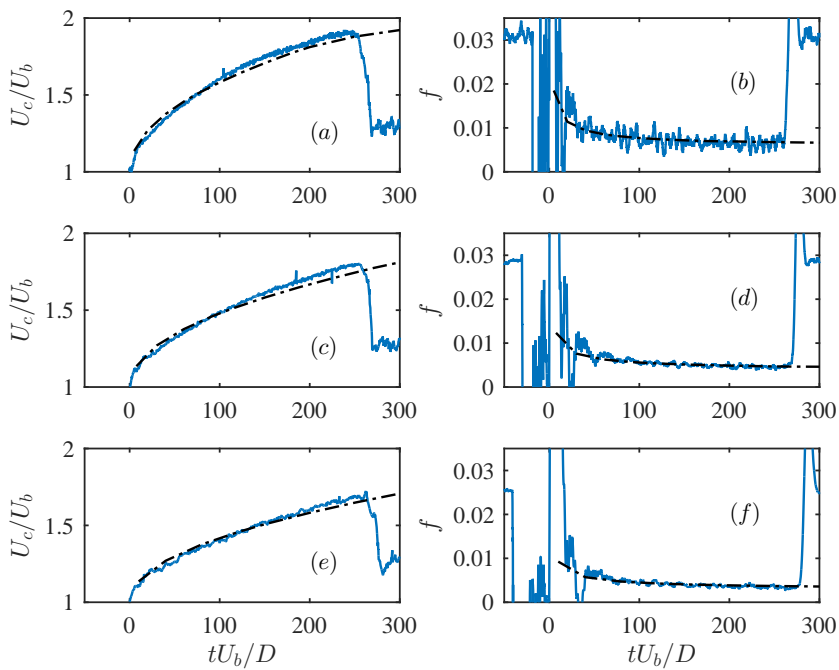}
\caption{Centreline velocity $U_c$ (left column) and friction factor $f$ (right column) as a function of time for relaminarising flow at $\Rey=10\,000$ (top row), $\Rey=15000$ (central row) and $\Rey=20\,000$ (bottom row). The wall stops at $t=0$. For the three cases $U_w=U_b$ and $s=s_c$. The dash-dotted line represents the development of a plug-like flow from a pipe entrance according to \citet{mohanty-1979}}
\label{fig.dvlp_vs_theroy_panel}
\end{figure}

The control strategy presented is effective in suppressing turbulence in the flow region perturbed by the wall and it allows us to study the flow development to the laminar state for a wide range of Reynolds numbers. In figure \ref{fig.dvlp_vs_theroy_panel} we compare the temporal evolution of the centreline velocity measured by LDV (left column) and the average friction factor $f=2\Delta p_1D /(\rho U_b^2 L_{taps})$ (right column), where $\rho$ is the water density and $L_{taps}$ is the distance between the two pressure taps.
The corresponding Reynolds numbers are $\Rey=10\,000$ (top row), $\Rey=15\,000$ (central row) and $\Rey=20\,000$ (bottom row). In this set of experiments we increased the length of the movable Perspex pipe to $L_{control}=385D$ to extended the duration of laminar flow. The evolution of $U_c$ and $f$ is compared with the development of a plug-like flow from a pipe entrance (dash-dotted line) according to the findings of \citet{mohanty-1979} and our data is in very good agreement with their prediction. To allow the comparison, we make the end of the wall motion ($t=0$) coincide with the entrance of the pipe and express the development in terms of advective time units. The friction factor adjusts rapidly to the one computed with the entrance problem model, thus suggesting that the transition from turbulent to laminar might happen rather quickly ($\lesssim 20 D/U_b$), and then the mean flow is nearly indistinguishable from a plug velocity profile evolving into parabolic. In addition, since the profile develops gradually from the wall, the friction factor approaches the laminar value much faster than the centreline velocity (cf. also figure \ref{fig.re=8000_u_evolution_panel} (b)). Hence, a substantial drag decrease is  obtained long before the laminar profile is fully developed.

The time available to observe the flow laminarising is constrained by the length of the control section and the shrinking of the laminar stretch which is being entrained by the surrounding turbulent flow (for entrainment rates of the turbulent fronts see e.g \citet{wygnanski-1973,Nishi2008,barkley-2015}). In particular, the turbulent front upstream of the laminar flow aggressively entrains it at rate that increases with Reynolds number.

\begin{figure}
\centering
\includegraphics[scale=.9]{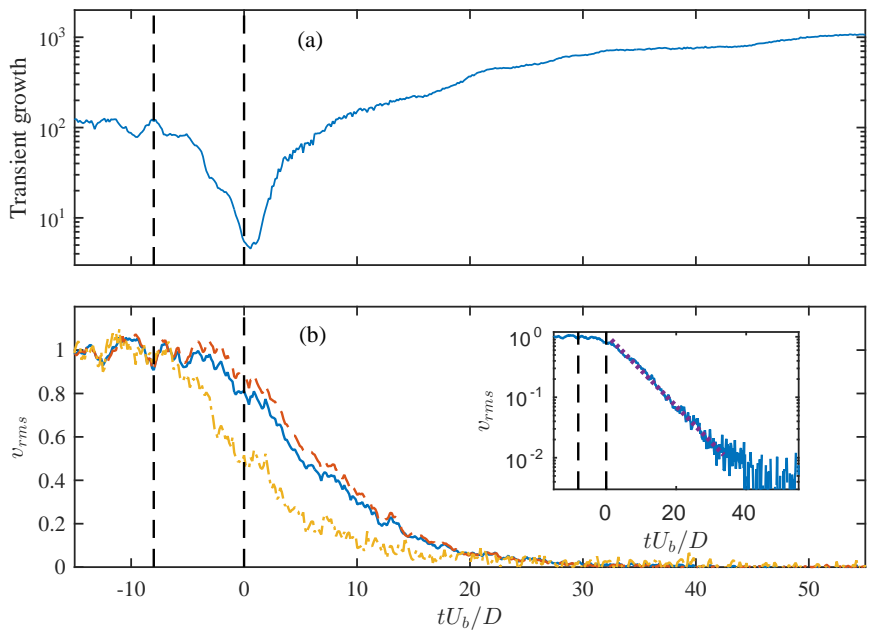}
\caption{Time evolution of (a), transient growth of the mean axial velocity and (b), mean wall-normal fluctuations over $0\leq y \leq D/2$ (solid line, blue online), $0.35D\leq y \leq D$ (dash-dotted line, yellow online), and $0 \leq y \leq 0.35D$ (dashed line, red online). The vertical dashed lines indicate the beginning and the end of the wall movement. The data is obtained by averaging 10 runs at $\Rey=5000$, with $U_w = U_b$ and $s=9D$. Inset, log-lin plot of the mean wall-normal fluctuations averaged for $0\leq y \leq D/2$ (solid line, blue online) and exponential fit (dotted line, purple online).}
\label{fig.re=5000_tg_vrms}
\end{figure}

We next focus on the physical mechanism responsible for the collapse of turbulence. As pointed out by \citet{kuhnen-2018}, flatter, more plug-like velocity profiles have lower levels of transient growth (TG). Consequently, the efficiency of the streak creation by streamwise vortices (lift-up mechanism) is reduced. While \citet{kuhnen-2018} demonstrated that relaminarising flows had lower levels of TG, we here consider the temporal evolution of the relaminarisation process in order to test the validity of this argument. As shown in figure \ref{fig.re=5000_tg_vrms} (a),
after the wall is set into motion TG starts to reduce and it does so at an increasing rate until the wall motion stops, which coincides with the minimum TG value reached. This sequence is in line with the development of the velocity profile (shown in figure \ref{fig.re=8000_u_evolution_panel} (a) for a different $\Rey$). The profile is not immediately flattened after the wall motion starts, but it first needs to adjust. As shown in figure \ref{fig.re=5000_tg_vrms} (a), during the interval of wall motion TG drops by approximately a factor of 20. We would expect that the fluctuation levels in the near wall region (i.e. where production takes place) react first and this is indeed the case, as demonstrated in figure \ref{fig.re=5000_tg_vrms} (b), where we show the time evolution of the  wall-normal velocity fluctuations
\begin{equation}
v_{rms} = \sqrt{\langle \left( V - \langle V \rangle \right)^2 \rangle},
\end{equation}
where $\langle.\rangle$ denotes averaging across the axial coordinate of the PIV window. The mean fluctuations in the wall region ($0.35D\leq y \leq D/2$, dash-dotted line, yellow online) closely follow the drop in TG while the average fluctuation level in the core region ($0 \leq y \leq 0.35D$, dashed line, red online) lags behind. In particular the strongest drop in the overall fluctuations (solid line, blue online) is assumed only somewhat after the minimum in TG has been reached, i.e. after the wall has stopped. While afterwards the TG level begins to rise, the value remains considerably lower than that of the average turbulent profile at this $\Rey$. Hence fluctuation levels keep decreasing. Eventually, when TG has regained its initial level ($tU_b/D\approx10$), the fluctuations, in particular in the near wall region, are very low and turbulence does not recover. Instead the profile becomes increasingly parabolic (cf. figure \ref{fig.re=8000_u_evolution_panel} (b)) and TG consequently continues to grow.

As shown in the inset of figure \ref{fig.re=5000_tg_vrms} (b), the decrease of the mean $v_{rms}$ that occurs after the wall motion is stopped can be approximated by an exponential. In this regime fluctuations drop by more than an order in magnitude in 20 advective time units. The exponential decay is also consistent with the findings of \citet{kuhnen-2018a} in the case of the relaminarising flow past an orifice plate obstacle. Qualitatively, above findings also agree with the recent work by \citet{marensi-2018}, who investigated numerically the robustness of optimal turbulence seeds in presence of a flat profile. In particular, the authors observed that a flatter base flow requires a greater initial disturbance energy and at the same time induces a smaller energy growth of the disturbances.

Revisiting the data shown in figure \ref{fig.criticalshiftlength_ubulk-vs-uopt_insert}, it appears that the shift length, and hence the time required to flatten the profile, scales linearly with the Reynolds number. In order to explain this trend, we look into the mechanism by which the axial velocity is progressively modified starting from the wall until it becomes flatter in the core region (see figure \ref{fig.re=8000_u_evolution_panel} (b)). To realise a plug profile, the new boundary condition established at the wall has to affect the entire flow up to the pipe centre. Assuming that the necessary profile modification occurs in viscous time scales, we propose that the adjustment up a to a distance $\delta$ from the wall requires a time
\begin{equation}
t \sim \frac{\delta^2}{\nu}.
\end{equation}
Substituting $\nu=U_bD/\Rey$ gives
\begin{equation}
t \sim \frac{\delta^2}{U_bD}\Rey,
\end{equation}
and hence, for a spread to the pipe centre $\delta\sim D/2$ we have
\begin{equation}
t \sim \frac{D}{U_b}\Rey.
\end{equation} 
The dimensionless time in advective time units $tU_b/D$ is then proportional to $\Rey$.

Owing to the advective nature of the flow, the linear growth of the necessary time for which the wall motion is active translates to a minimum streamwise length  that needs to be exposed to the changed boundary conditions. This observation also explains why a related control strategy where the flow is accelerated  by streamwise fluid injection at a fixed location only works for a finite Reynolds number range \citep{kuhnen-2018a}.

\section{Conclusions}\label{sec.conclusions}
We demonstrated that upon an abrupt acceleration of the near wall fluid, the transient growth level of the overall flow is strongly suppressed and subsequently turbulent (wall normal) fluctuation levels drop exponentially. While at low $\Rey$ ($\approx5000$) arbitrarily large wall speeds lead to realminarisation, at higher $\Rey$ only wall speeds close to the bulk flow speed lead to a decay of turbulence. Moreover the wall motion required to accelerate the near wall fluid has to act for a minimum time in order to create the desired plug flow, because the velocity profile adjusts viscously from the boundaries. This requirement severely limits the applicability of such relaminarisation schemes that affect the flow only at the boundaries, since due to the advective nature of the flow it effectively means that the control has to act over a minimum distance in the streamwise direction which increases linearly with Re. It should be noted that such limitations do not apply if the profile can be adjusted via a volume force as shown by a numerical forcing scheme by \citet{kuhnen-2018}. As those authors demonstrated relaminarisation under such conditions can even be achieved at Re as large as 100\,000.
\\
\\
We acknowledge the European Research Council under the European Union's Seventh Framework Programme (FP/2007-2013)/ERC Grant Agreement 306589, the European Research Council (ERC) under the European Union's Horizon 2020 research and innovation programme (grant agreement no. 737549). We thank M. Parvulescu for carrying out several measurement campaigns.

\bibliographystyle{jfm}
\bibliography{Literatur_JK,ds}

\end{document}